\documentclass[conference,10pt]{IEEEtran}
\IEEEoverridecommandlockouts
\usepackage{cite}
\usepackage{amsmath,amssymb,amsfonts}
\usepackage{algorithmic}
\usepackage{graphicx}
\usepackage{textcomp}
\usepackage{xcolor}
\usepackage{color} 
\def\BibTeX{{\rm B\kern-.05em{\sc i\kern-.025em b}\kern-.08em
    T\kern-.1667em\lower.7ex\hbox{E}\kern-.125emX}}
\usepackage{amsfonts}
\usepackage{tikz}
\usepackage{enumitem}
\usepackage{tabularx}
\usepackage{array}
\usepackage{multirow}
\usepackage{booktabs} 
\usepackage{caption}

\definecolor{ACMPurple}{cmyk}{0.55,1,0,0.15}
\definecolor{ACMBlue}{cmyk}{1,0.1,0,0.1}
\definecolor{greenfont}{RGB}{0,128,0}

\usepackage{hyperref}
\hypersetup{
    colorlinks,
    linkcolor=ACMPurple,
    filecolor=ACMPurple,      
    urlcolor=ACMBlue,
    citecolor=ACMPurple,
}

\begin{document}

\title{FaaSRCA:
Full Lifecycle Root Cause Analysis for Serverless Applications
}

\author{\IEEEauthorblockN{Jin Huang$^{1}$, Pengfei Chen$^{1*}$, Guangba Yu$^{1}$, Yilun Wang$^{2}$, Haiyu Huang$^{1}$, Zilong He$^{1}$}
\IEEEauthorblockA{$^{1}$School of Computer Science and Engineering, Sun Yat-sen University, Guangzhou, China \\
$^{2}$School of Systems Science and Engineering, Sun Yat-sen University, Guangzhou, China \\
Email: \{huangj323, yugb5, wangylun6, huanghy, hezlong\}@mail2.sysu.edu.cn, $^{*}$chenpf7@mail.sysu.edu.cn}
}
\maketitle

\begin{abstract}
Serverless becomes popular as a novel computing paradigms for cloud native services. However, the complexity and dynamic nature of serverless applications present significant challenges to ensure system availability and performance. There are many root cause analysis (RCA) methods for microservice systems, but they are not suitable for precise modeling serverless applications. This is because: (1) Compared to microservice, serverless applications exhibit a highly dynamic nature. They have short lifecycle and only generate instantaneous pulse-like data, lacking long-term continuous information. (2) Existing methods solely focus on analyzing the running stage and overlook other stages, failing to encompass the entire lifecycle of serverless applications. To address these limitations, we propose \textit{FaaSRCA}, a full lifecycle root cause analysis method for serverless applications. It integrates multi-modal observability data generated from platform and application side by using \textit{Global Call Graph}. We train a Graph Attention Network (GAT) based graph auto-encoder to compute reconstruction scores for the nodes in global call graph. Based on the scores, we determine the root cause at the granularity of the lifecycle stage of serverless functions. We conduct experimental evaluations on two serverless benchmarks, the results show that \textit{FaaSRCA} outperforms other baseline methods with a top-k precision improvement ranging from 21.25\% to 81.63\%.
\end{abstract}


\section{Introduction}
Serverless, or Functions-as-a-Service (FaaS) has now become a popular application deployment and execution model\cite{fouladi2019laptop, Amazon2020, Sam2021}. Extensive research has been conducted in the field of serverless  \cite{alpernas2021cloud, datta2022alastor, stojkovic2023mxfaas, yu2023faasdeliver, WangCF22}. Serverless architecture decomposes applications into small, independent functions that communicate with each other. These functions run in temporary containers, which are dynamically created and scheduled in response to events such as web requests, file system reads/writes, etc. Serverless can scale elastically based on incoming requests, enabling resource provisioning aligned with demands. In contrast to traditional methods, serverless allows users to easily add resources as system load increases and remove them when load decreases. It can even deactivate all resources and scale down to zero, avoiding the cost of idle resources.

Despite above benefits, serverless applications still encounter challenges. Serverless functions are deployed across multiple hosts and are frequently scaled up or down to adapt to changing business requirements. In such a highly dynamic environment, serverless functions are susceptible to failures stemming from various reasons. To ensure the availability and performance of serverless systems, it is crucial to quickly and accurately localize the root cause of issues. Root cause analysis (RCA) pinpoints the specific cause responsible when an anomaly is detected \cite{hou2021diagnosing, Eadro23, wu2020microrca, zhang2021cloudrca}. However, we identify some limitations in these existing RCA approaches.
\begin{enumerate}[itemsep=0pt, topsep=0pt]
  \item \textbf{Lack of consideration for the highly dynamic nature of serverless applications.} Different from microservices, the lifecycle of a serverless application is significantly shorter \cite{kaffes2019centralized}. Serverless functions typically exist for a duration ranging from hundreds of milliseconds to minutes. Once their executions are completed, they are terminated instead of running continuously. As a result, they generate instantaneous pulse-like data rather than long-term continuous data. This distinction prevents us from accurately modeling them in the same manner as microservices.
  
  \item \textbf{Neglect the entire lifecycle of serverless applications.} Current RCA methods only focus on the running stage and neglect other stages, thus lacking comprehensive coverage of the entire lifecycle of applications. They only use observability data from the application side to locate the root cause, while ignoring the platform side. Platform side refers to the underlying infrastructure where serverless functions are deployed, while application side refers to the instances of functions. However, when the system platform fails, serverless functions will suffer errors that can only be detected with platform side data.
  
  \item \textbf{Only use single modal data to diagnose problems.} Most existing RCA methods primarily rely on single modal data, which limits their ability to provide comprehensive insights into the system's state. These methods fail to reveal a wide range of anomalies, resulting in a lack of accurate root cause analysis. For instance, trace-based RCA methods \cite{liu2020unsupervised, yu2021microrank, yu2021tracerank} only offer relatively coarse-grained service-instance-level root causes due to limited traces granularity. They struggle to identify local abnormal behaviors within services, such as resource exhaustion, thereby missing out on these failures.
  
  
\end{enumerate}



To overcome above limitations, we propose \textit{FaaSRCA}, a full lifecycle RCA approach for serverless applications, which integrates and analyzes multi-modal observable data from both the platform and application side. \textit{FaaSRCA} not only allows us to identify the specific functions (e.g., \textit{func\_starter}, \textit{func\_load}, etc.), but also enables the analysis of the stage (e.g., \textit{creation}, \textit{execution}) at which the faults occur. It takes into account the unique characteristics of FaaS and the lifecycle on the platform side, effectively addressing serverless RCA issues.

To be specific, \textit{FaaSRCA} consists of following parts. (1) \textit{Data Preparation.} Initially, we gather observability data from both the platform and application side. (2) \textit{Data Transformation.} We utilize Bidirectional Encoder Representation from Transformers (BERT) \cite{devlin2019bert} to encode logs. Additionally, we employ an embedding method to encode metrics and traces. Finally, we fuse these encoded representations to obtain the unified vector. (3) \textit{Graph Construction.} We construct the \textit{Global Call Graph} by using platform and application side traces topology. (4) \textit{Model Training.} Graph Attention Network\cite{velivckovic2017graph} (GAT) is utilized as a graph auto-encoder to compute the reconstruction scores of the nodes. Next, we calculate the score distribution for each node to obtain the normal pattern of the graph. (5) \textit{Root cause analysis.} For serverless applications requiring root cause analysis, we compare their faulty global call graphs with the corresponding normal ones. By assessing the degree of deviation of each node from its normal state,  we rank the nodes accordingly. The top-ranked nodes are identified as potential root causes. The implementation of \textit{FaaSRCA} is publicly available at \cite{Anonymous-FaaSRCA-1524}.

In summary, our contributions are as follows.
\begin{itemize}[itemsep=0pt, topsep=3pt]
   \item We identify limitations when using existing root cause analysis methods in serverless. To effectively model the behavior of serverless applications, we introduce \textit{Global Call Graph} to integrate multi-modal observability data from the platform and application side.
  \item We propose \textit{FaaSRCA}, an unsupervised method which enables full lifecycle root cause analysis for serverless applications. It can pinpoint the root cause to specific lifecycle stages of certain serverless functions. 
  \item We conduct extensive experiments on two serverless benchmark datasets to validate the effectiveness of \textit{FaaSRCA}. The results indicate that it outperforms all compared approaches, exhibiting an improvement of 21.25\% in top-k precision compared to the best baseline method.
\end{itemize}

\section{Background}
\subsection{Serverless}
Serverless architecture has gained popularity due to its ability to create elastic and scalable applications. It offers advantages such as demand-driven resource allocation and fine-grained billing based on usage. As a result, serverless architecture has been widely adopted in various domains, including e-commerce applications \cite{Sam2021}, blog services \cite{RealWorld2019}, distributed builds \cite{fouladi2019laptop}, and more. The fundamental units of serverless applications are functions that run in transient containers. However, the use of temporary containers also imposes some limitations on serverless applications \cite{alpernas2021cloud, hellerstein2018serverless}. The lack of long-lived instances means that serverless functions don't have long-term persistent data. Additionally, the frequent creation and destruction of function instances make them more vulnerable to the influence of system platform.

\subsection{Multi-modal observability data}
Observability data typically refers to three types of data (i.e., metrics, logs, and traces). Metrics are numerical values that provide insights into the system behavior. They provide information about system performance and its current state. Logs are textual records that capture status and execution operations of the system. They can report exceptions and help understand various events occurring in the system. Traces consist of a collection of spans and the causal relationships between them. They capture the call paths that requests traverse within the system, providing insights into the interactions among different system components. Besides, they record the execution duration of each component along the path.

\section{Motivation}
\label{sec:motivation}

\subsection{Are the RCA methods for microservice suitable for serverless application with highly dynamic nature?}
\label{subsec:short_lifecycle}

\begin{figure}[t]
    \centering
    \includegraphics[width=0.48\textwidth]{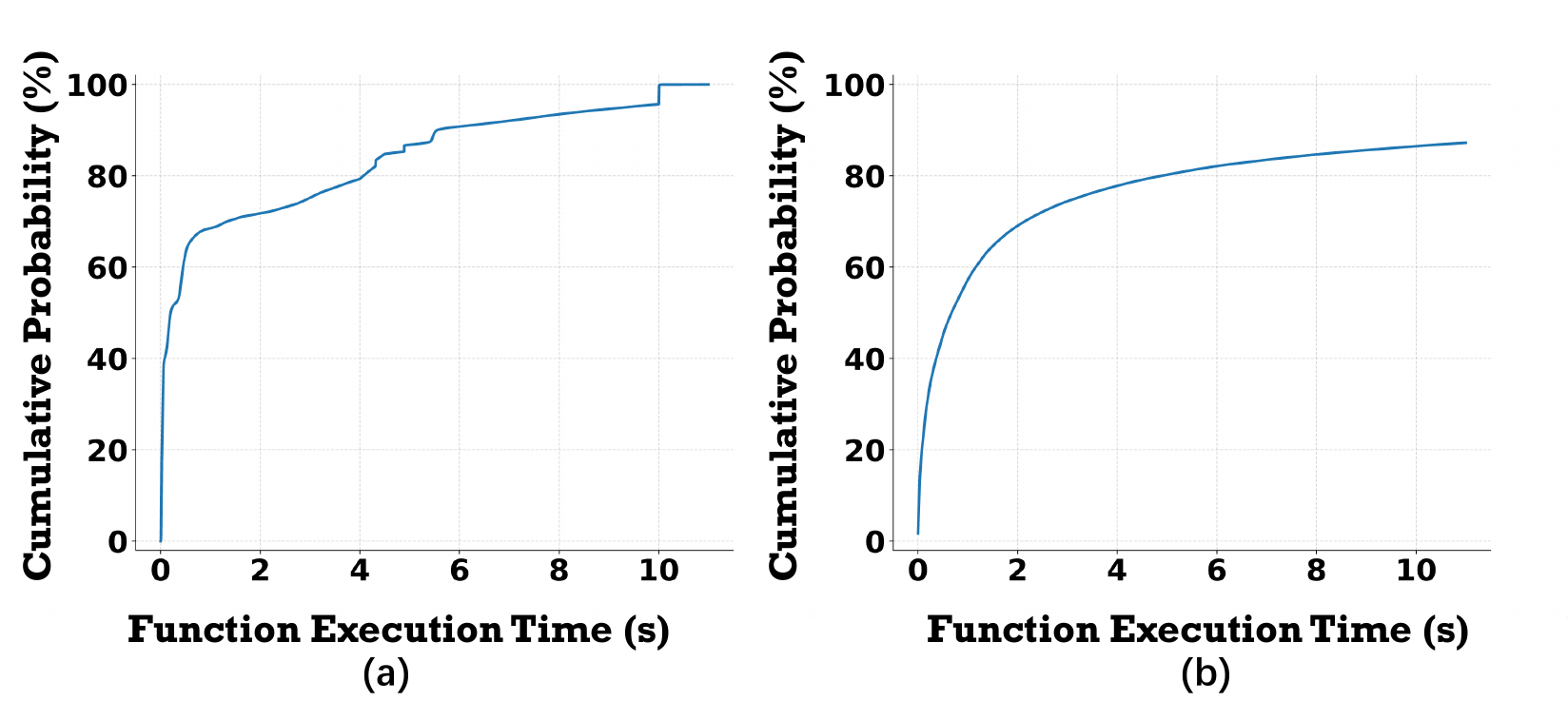}
    \captionsetup{font=normal}
    \caption{(a) Functions execution time in Serverless TrainTicket. (b) Functions execution time in Microsoft Azure.}
    \vspace{-0.2in}
    \label{fig: Function_execution}
\end{figure}

Serverless functions, unlike long-term microservices, have transient lifecycle and run for short duration within temporary containers. We use data collected from Serverless TrainTicket \cite{ServerlessTrainTicket2023} as an example. Fig.~\ref{fig: Function_execution}(a) shows that out of the 800,368 serverless functions executed in the past, 726,590 have execution time ranging from 0 to 6 seconds, accounting for 90.78\% of the total. And all functions have execution time of less than 11 seconds. Moreover, Microsoft Azure released a production trace of its serverless offering \cite{ShahradFGCBCLTR20}, comprising data on 445M function invocations across their infrastructure over 14 days. The median function execution time is just 600 milliseconds, as shown in Fig.~\ref{fig: Function_execution}(b), confirming the short-lived nature of serverless functions from a practical system perspective.

\begin{figure}[h]
    \centering
    \includegraphics[width=0.43\textwidth]{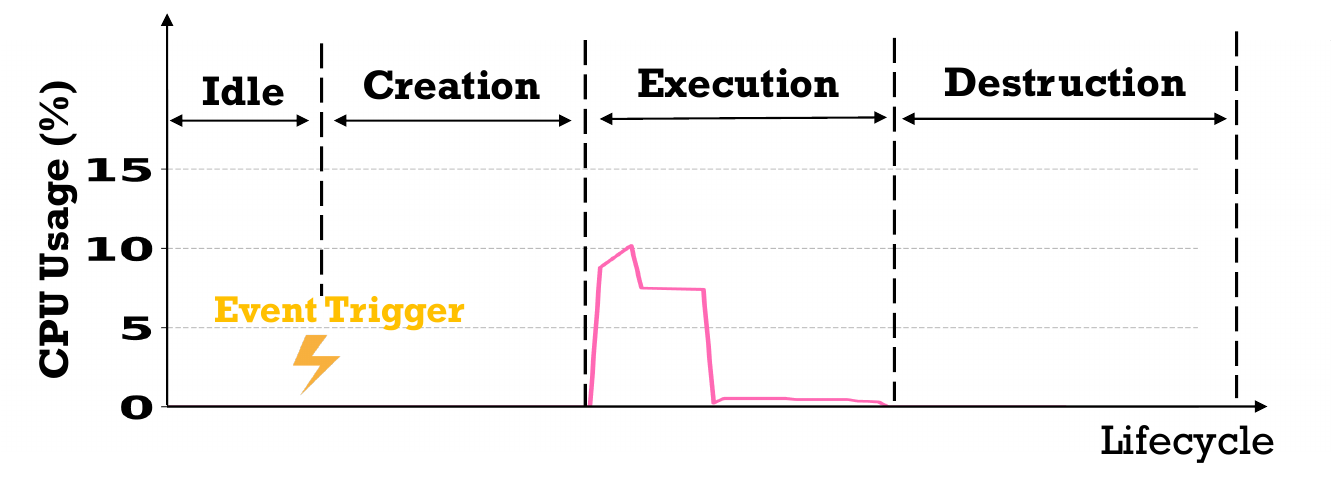}
    \captionsetup{font=normal}
    \vspace{-0.1in}
    \caption{CPU usage variation in serverless functions’ lifecycle.}
    \vspace{-0.1in}
    \label{fig:burst}
\end{figure}

On the other hand, services in microservice system continue to occupy resources even if no requests are issued. However, in serverless, if there are no incoming requests, the function instances will shrink to zero and no longer consume any system resources. Fig.~\ref{fig:burst} shows the variation of CPU utilization over time in the lifecycle of a serverless function. It illustrates that serverless functions are triggered by events and have short lifecycle, preventing them from exhibiting stable behavior. Consequently, their data changes demonstrate a pulse-like, discontinuous, and non-differentiable nature. Sequence-based methods are not adapted to the characteristics of serverless, may struggle to model and analyze serverless functions.

\begin{figure}[h]
    \centering
    \includegraphics[width=0.45\textwidth]{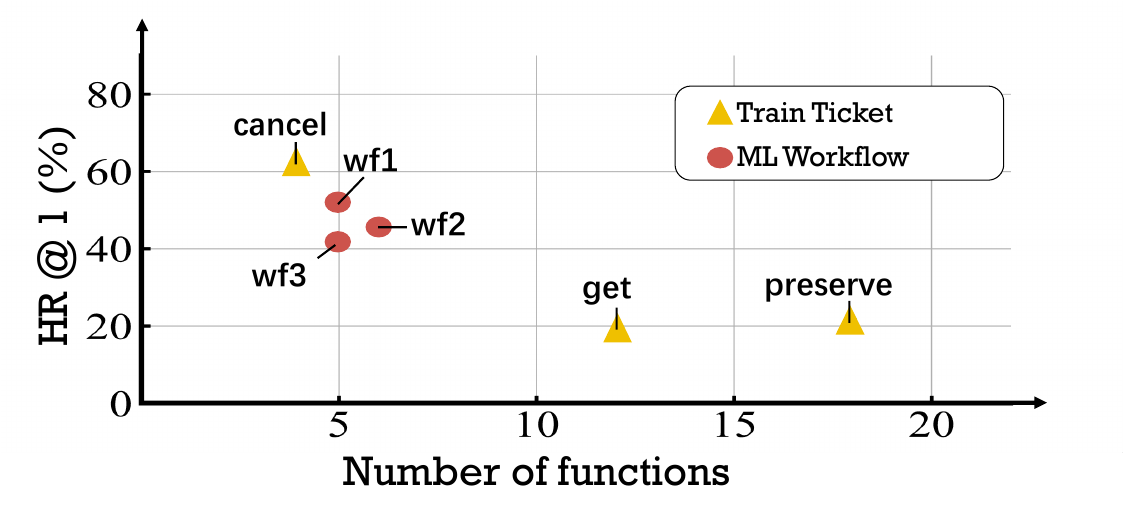}
    \captionsetup{font=normal}
    \caption{Evaluation results of Eadro on two serverless datasets.}
    \vspace{-0.1in}
    \label{fig:eadro}
\end{figure}

To illustrate this point, we use the advanced microservice RCA method Eadro\cite{Eadro23}, to conduct experiments on two serverless datasets -- Serverless TrainTicket \cite{ServerlessTrainTicket2023} and ML Workflow \cite{wang2022enhancing}. Since our focus is on the performance of sequence-based methods with transient data, we specifically analyze cases where the root cause is related to metrics, and utilize metric sequences of each function as input. Fig.~\ref{fig:eadro} shows the evaluation results for different types of requests (\textit{cancel}, \textit{preserve}, \textit{wf1}, etc.) in the datasets. It shows that Eadro's performance in top-1 precision (HR@1: probability of the first predictive element being the root cause) is relatively low. Moreover, its effectiveness further deteriorates as the number of functions included in the request increases. These results suggest that RCA methods designed for microservices do not perform well for serverless applications, due to the absence of continuous data that can be available for analysis. 

\subsection{Is it sufficient to focus solely on application side data?}
\label{subsec:platform_side}

\begin{figure}[t]
    \centering
    \includegraphics[width=0.45\textwidth]{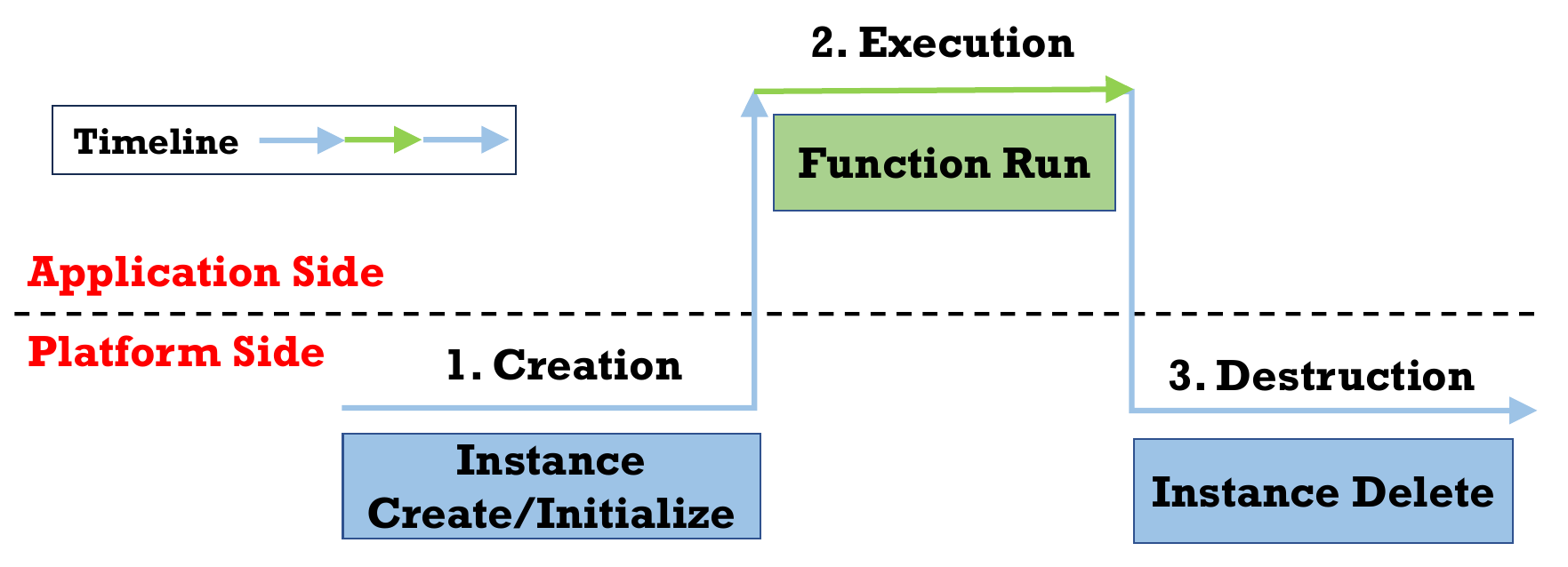}
    \captionsetup{font=normal}
    \caption{The lifecycle stages of serverless functions.}
    \vspace{-0.2in}
    \label{fig:lifecycle}
\end{figure}

The lifecycle of serverless functions is shown in Fig.~\ref{fig:lifecycle}, encompassing the stages of creation, execution, and destruction. Creation and destruction stages belong to the platform side, while execution stage belongs to application side. From the client's perspective, client side factors primarily affect the execution of the functions on the application side. However, if the platform infrastructure fails, serverless functions may also encounter errors during creation stage. For instance, network loss can cause image pull failure, result in the inability to create instances, but existing methods 
miss these faults.

Tim et.al.\cite{goodwin2023goes} analyzed the reports published on the public Github issue of the open source serverless platform Knative \cite{Knative2023}. They discovered a total of 103 issues occurring on the platform side, including 32 bugs related to interactions with Kubernetes, 40 errors related to components health/readiness status, and so on. If we solely collect data from the application side, we risk overlooking platform side anomalies. To locate root cause across the full lifecycle of serverless functions, it is crucial to incorporate platform behavior into analysis.

\subsection{Can accurate root cause analysis be achieved only by relying on single modal data?}
\label{subsec:multi_modal}

Traces focus on recording the interactions between different services and provide valuable information, such as latency and system topology. However, they have limitations in capturing behavior within individual services and identifying local anomalies. For example, if a Kubernetes pod experiences a failure, detailed information about its internal state can be obtained from logs rather than traces. Logs can report internal anomalies in services by presenting abnormal patterns, such as the presence of error keywords. But in certain abnormal scenarios, logs may not provide key information directly related to the root cause.
In cases of CPU contention issues, we can detect abnormally high CPU utilization, but clear abnormal patterns may not be evident in the logs. Metrics capture fine-grained information about resource usage and service performance, providing additional insights to identify failures. However, metrics are generated at the local level by each service, lacking consideration for inter-service dependencies. Relying solely on metrics can lead to a loss of global context, degrading the performance of RCA methods.

Our experimental findings in \S\ref{subsec:exp_multi_modal} further support our idea. Compared to employing all multi-modal data, the absence of a specific data source leads to lower top-k precision for the method, ranging from 25.02\% to 39.20\%. In conclusion, relying on single modal data may not reveal all abnormal situations, and may overlook valuable information contained in other data sources. To fully comprehend the system, it is necessary to integrate multi-modal data.

\section{Approach}
\begin{figure*}[h]
    \centering
    \includegraphics[width=0.8\textwidth]{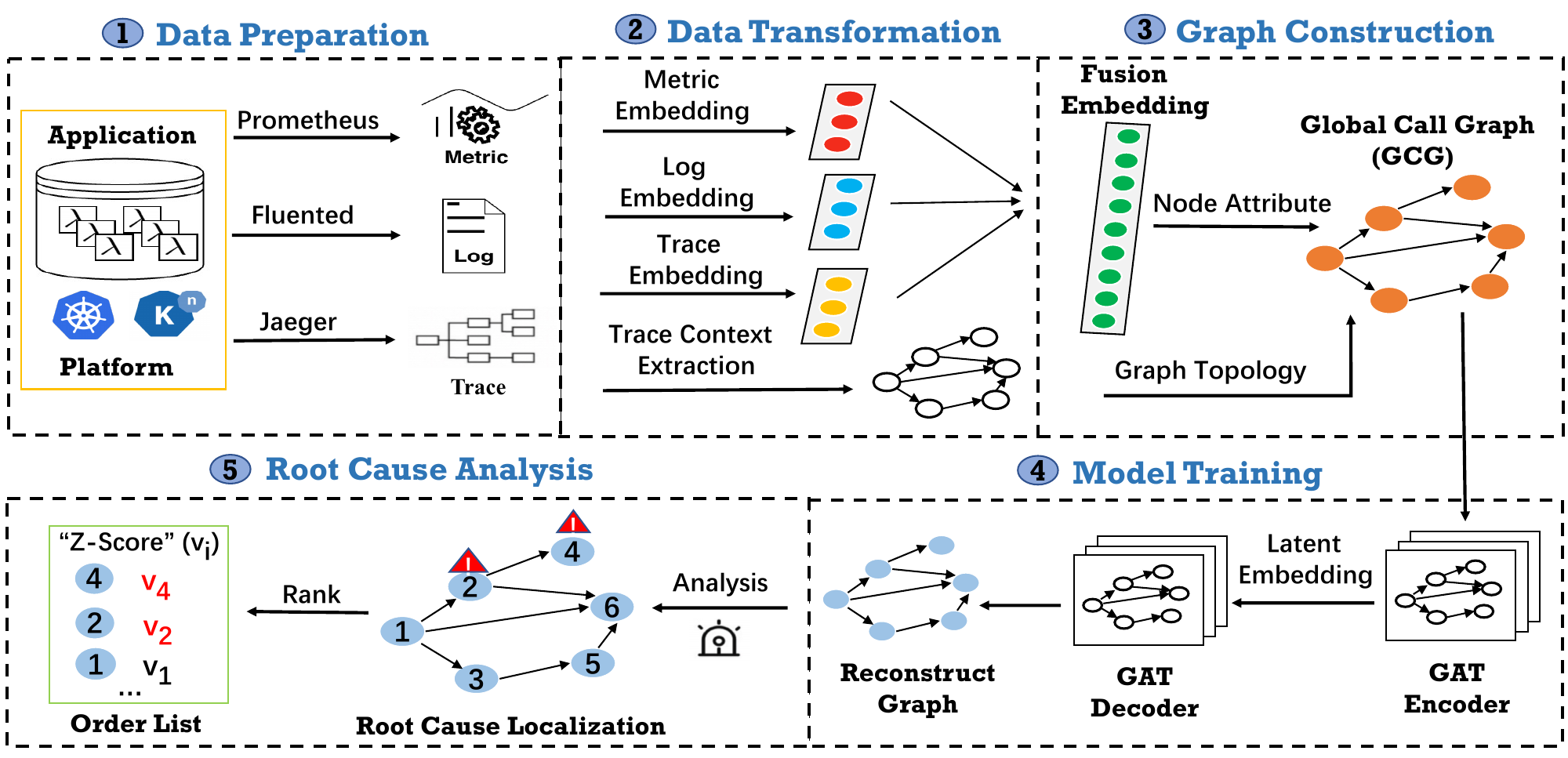}
    \captionsetup{font=normal}
    \caption{Overall architecture of \textbf{\textit{FaaSRCA}.}}
    \vspace{-0.2in}
    \label{fig:FaaSRCA_overview}
\end{figure*}

In this paper, we introduce \textit{FaaSRCA}, a full lifecycle RCA method specifically designed for serverless applications. Fig.~\ref{fig:FaaSRCA_overview} shows the overview of \textit{FaaSRCA}, which consists of five parts. Specifically, to accommodate the high dynamism and short lifecycle of serverless applications proposed in \S\ref{subsec:short_lifecycle}, we provide detailed introductions in \textbf{\textit{Data Transformation}} on how to handle their transient data. To consider the entire lifecycle of serverless applications (\S\ref{subsec:platform_side}), in \textbf{\textit{Data Preparation}} we discuss how to collect observability data from both the platform and application side. To leverage the multi-modal data for accurate diagnosis (\S\ref{subsec:multi_modal}), we explain how to integrate these data to form a global call graph in \textbf{\textit{Graph Construction}}, and we train a GAT based graph auto-encoder in \textbf{\textit{Model Training}}. Lastly, we compute reconstruction scores of the nodes for root cause localization in \textbf{\textit{Root Cause Analysis}}. 

\subsection{Data Preparation}
\label{subsec:data_preparation}
\textbf{Metrics.} Metrics capture the internal behavior of serverless functions, their abnormal changes often indicate anomalies. Our focus is on system-level metrics, such as memory and CPU usage, which offer insights into the local root causes.

\textbf{Logs.} Since serverless applications typically utilize Kubernetes as the platform, we collect two types of log provided by Kubernetes: \textit{``Audit''} and \textit{``Event''}. \textit{Audit} records the access to Kubernetes cluster resources, while \textit{Event} captures information about the components of Kubernetes. These logs allow us to monitor activities within the platform. By default only the \textit{Event} from the last past hour are stored in Kubernetes. To overcome this limitation, we employ Elasticsearch \cite{Elastic2023} as the storage back end to continuously store these logs.

\textbf{Traces.} In the Kubernetes platform, tracing mechanisms between components are not inherently built-in due to its asynchronous and declarative nature. As a result, traditional distributed trace models are not suitable for Kubernetes. However, we have observed that there are implicit causal relationships among Kubernetes components. For instance, the creation of serverless function container instances involves Kubernetes components such as deployment, replicaset, and pod, and their interaction can be represented in a single trace. Inspired by this, we establish connections among Kubernetes components based on their ownership relationships and abstract the process of instance creation as a trace. By organizing each component undergoing operations as a span and linking them together, we create a trace that captures the behavior of multiple components from a global perspective.

\begin{figure}[h]
    \centering
    \includegraphics[width=0.5\textwidth]{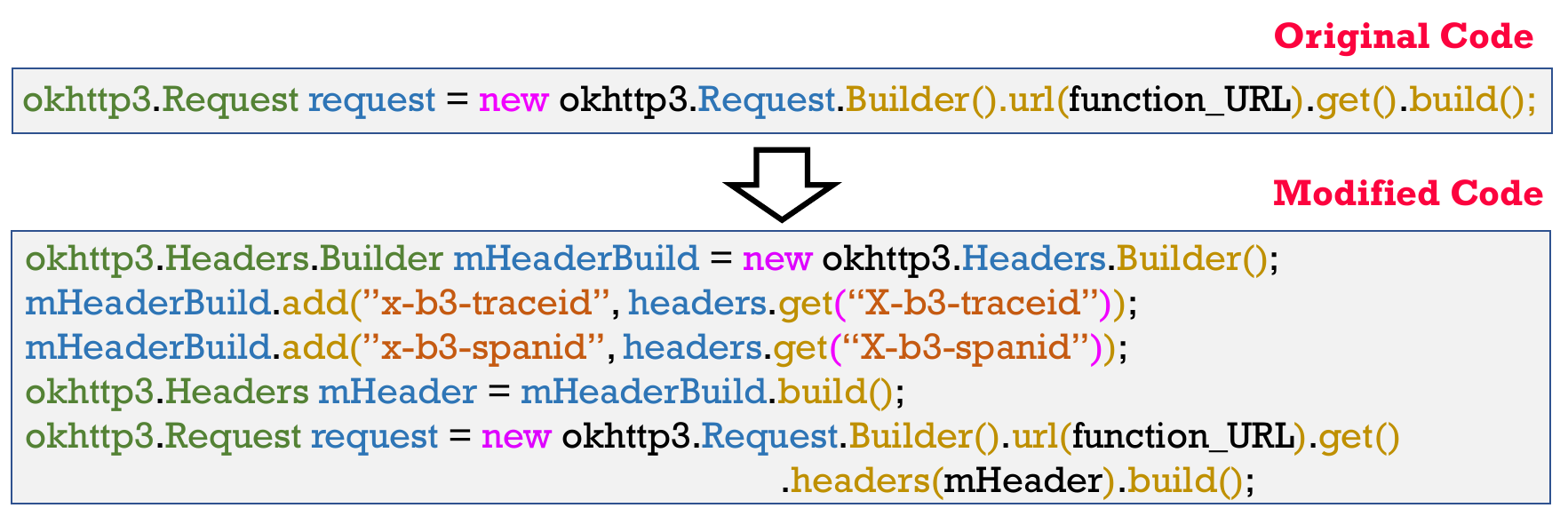}
    \captionsetup{font=normal}
    \caption{Example of propagating trace context in function code.}
    \vspace{-0.05in}
    \label{fig:opentelemetry_trace}
\end{figure}

On the application side, if the dataset lacks built-in trace recording capabilities, we adopt the unified standard of OpenTelemetry \cite{Opentelemetry2023}. To implement trace injection and collection in them, we utilize tools like OpenTelemetry SDK and Jaeger \cite{Jaeger2023}. For example, in the source code of the functions, we can modify the relevant lines shown in Fig.~\ref{fig:opentelemetry_trace} to construct the HTTP header. This header includes the trace and span IDs, which are propagated in subsequent calls to different functions. Additionally, to form a complete trace, we incorporate trace collection on the gateway (e.g., Istio \cite{Istio2023}) of the serverless platform. By leveraging traces context, we capture the invocation latency of functions and their topology structure.

\subsection{Data Transformation}
\subsubsection{Log Embedding}
To extract the constant part of logs, we employ log parsing methods. Specifically, we utilize Drain \cite{he2017drain} to parse the original logs and generate structured format log templates. Once we obtain these templates, we split each sentence into individual words using common delimiters such as spaces and commas. Then, we convert all words to lowercase and remove any non-verbal symbols like punctuation and number. At last, we filter out stop words from them. 

To generate semantic representations for each log, we employ log embedding techniques. Using the log templates as basis, we utilize the transformer based pre-trained language model BERT \cite{devlin2019bert}. BERT offers extensive word embeddings to encode words. We use the BERT-Base model provided by \cite {wolf2020transformers}, which includes 12 layers of transformer encoders, each with 768 dimensional hidden units. By utilizing multiple layers of attention mechanisms and incorporating contextual information, the model aggregates the hidden states of all words to generate a global representation. For each sentence, we can generate a 768 dimensional vector $\textbf{\textit{E}}_{sentence}$ which captures the semantics of the sentence.

Once the embedding for each log is obtained, we will use a unified vector to represent the sequence composed of multiple log sentences. To achieve this, we first remove duplicate logs from the sequence. Then for each log, we convert it into a corresponding vector $\textbf{\textit{E}}_{sentence}$, and employ sum pooling to obtain the vector $\textbf{\textit{E}}_{log}$ for the entire log sequence. It should be noted that \textit{Audit} and \textit{Event} logs possess distinct characteristics, thus requiring separate calculations for them.

\subsubsection{Metrics \& Traces Embedding}
To address the challenge of lacking long-term continuous data in highly dynamic and short-lived serverless applications, we leverage their instantaneous data (consisting of single scalar values) for root cause analysis. However, since the embedded logs have high dimension, if the metrics features correspond to only a single dimension, the displayed information may be limited and potentially overlooked in the representation. Li et.al. \cite{li2018time} proposed a time-dependent event representation method. Inspired by it, we project the metrics onto a high-dimensional vector space and expand them. 
To begin, we project the scalar values $\textit{x}$ (one-dimensional feature) of the metrics onto a $p$-dimensional vector space. Subsequently, we multiply it with a randomly initialized weight vector $\textbf{\textit{W}}$ and add a bias vector $\textbf{\textit{b}}$. After the linear transformation, we apply $softmax(\cdot)$ function, which rescales the tensor. This ensures that each element is within the range of (0, 1) and that the elements along the selected dimension sum up to 1.
\begin{equation}
\textit{s}=softmax(\textbf{\textit{W}}x+\textbf{\textit{b}}),\quad\textbf{\textit{W}}\in \mathbb {R}^{p}, \textbf{\textit{b}}\in \mathbb {R}^{p}.
\end{equation}
Lastly, we weight all rows in randomly initialized embedding matrix $\textbf{\textit{E}}_{s}$ using the values in $\textit{s}$, obtain metric embedding $\textbf{\textit{E}}_{metric}$, which have the same dimension as $\textbf{\textit{E}}_{log}$. 
\begin{equation}
\textbf{\textit{E}}_{metric}=\textit{s}\textbf{\textit{E}}_{s},\quad \textbf{\textit{E}}_{s}\in \mathbb {R}^{p\times d}.
\end{equation}

Furthermore, to detect performance issues such as extended creation time due to network congestion or prolonged execution time caused by code defects, it is essential to analyze trace latency. On the platform side, we calculate the time difference between the start and end logs within each Kubernetes component as trace latency, which represents the time required for them to perform their operations. On the application side, we consider duration of the function execution as the latency. Performance issues can be identified by analyzing changes in latency, so we employ latency analysis in a similar way to embed metrics, obtain trace embedding $\textbf{\textit{E}}_{trace}$.
\subsection{Graph Construction}
\textit{FaaSRCA} utilizes an attribute graph named \textit{Global Call Graph}, denoted as $G = \{V, E, X\}$, to construct the complete lifecycle of serverless application requests. $V$ comprises two sets of nodes: $V_{p}$ for platform side nodes, including Kubernetes components involved in creating function instances, and $V_{a}$ for application side nodes, representing the function instances. We are able to pinpoint the root cause to the granularity of individual components. These nodes' granularity aligns with the logic of serverless functions being created and providing service, and it is intuitive, without the need for additional expertise to further partition the analysis.

$E=\{E_{p},E_{a}\}$ denotes the collection of edges. An edge $\vec e_{m,n}=(v_{m},v_{n}) \in E$ indicates a directed connection from node $v_m$ to node $v_n$. Traces are utilized to represent these connections between nodes in the graph. On the platform side, the logical relationships between Kubernetes components are treated as edges $E_{p}$. On the application side, the invocation relationships between different functions serve as edges $E_{a}$. Based on the common unique name -- the service name on the application side, which is also the pod name on the platform side, we merge the traces from both sides to construct a comprehensive topology of the global call graph, representing the full lifecycle of FaaS. At last, we integrate multi-modal data from the platform and application side, covering the entire lifecycle of serverless applications.

$X \in \mathbb {R}^{|V|\times d}$ is the node attribute matrix, each row $x_{v}$ in $X$ corresponds to the vector of a node $v\in V$, where $d$ is the dimension of vector. To leverage the diverse modalities of data, we employ the intermediate fusion approach which combines representations from different modalities. Firstly, we individually embed logs, metrics, and traces into vectors. Then, we concatenate and fuse them together to obtain comprehensive representations for each node.

\begin{figure}[t]
    \centering
    \includegraphics[width=0.4\textwidth]{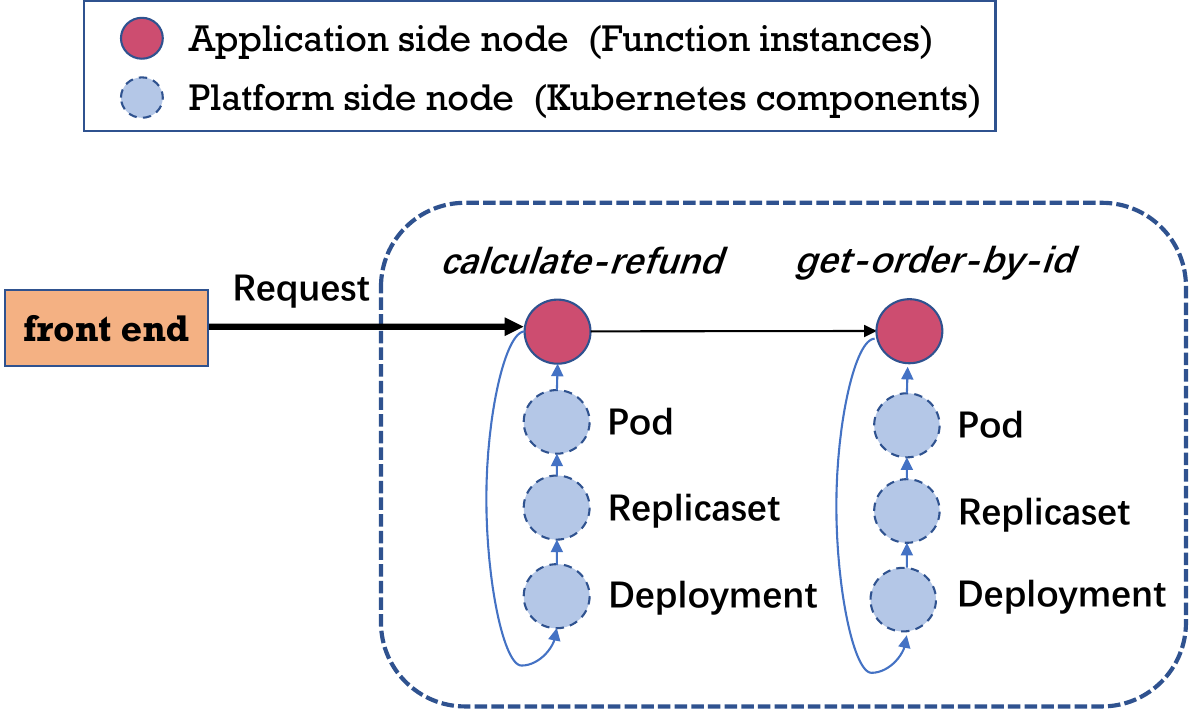}
    \captionsetup{font=normal}
    \caption{Visualization of \textit{Global Call Graph}.}
    \vspace{-0.2in}
    \label{fig:Global_Call_Graph}
\end{figure}

In Fig.~\ref{fig:Global_Call_Graph}, we demonstrate the generation of a \textit{Global Call Graph}. When a calculate request is issued by the front end, it first follows the logical relationship on the platform side, traversing through the Kubernetes components (deployment, replicaset, pod). Once the instance of the \textit{``calculate-refund''} function is successfully created, the request transits to the application side and propagates through the call relationship between functions. Afterwards, the request enters the platform side of the corresponding called function \textit{``get-order-by-id''} to invoke it. This process continues until the request is completed.

\subsection{Model Training}
To leverage the rich information in the graph, we employ Graph Neural Network (GNN) in our method. Inspired by previous work \cite{ding2019deep}, we utilize Graph Auto Encoder (GAE) to detect abnormal nodes. The learning process involves minimizing a cost function with an input dataset $\textbf{\textit{X}}$.

\begin{equation}
 min\mathbb{E}[dist(\textbf{\textit{X}}, Dec(Enc(\textbf{\textit{X}}))],
\end{equation}
where $dist(\cdot,\cdot)$ is a predefined distance metric, $Enc(\cdot)$ is the encoder, and $Dec(\cdot)$ is the decoder. Due to the potential issue of excessive smoothing in the simple aggregation operation of Graph Convolution Network (GCN) \cite{KipfW17}, it becomes challenging to distinguish between abnormal and normal nodes. Hence, we adopt stacked layers of Graph Attention Network (GAT) \cite{velivckovic2017graph} as the encoder to learn embedding vectors for the nodes. And we use stacked layers of GAT as decoder to reconstruct the node attribute matrix $\textbf{\textit{X}}$ from the latent embeddings.

The core idea of GAT is to introduce attention mechanisms among nodes. It computes weights for each neighboring node to learn the dependency relationships between nodes and their respective neighbors. Specifically, for each node $i$ in the graph, GAT first multiplies its original features by a trainable weight matrix to obtain its representation. It is then passed through a LeakyReLU non-linear activation function. The attention coefficients, which indicate the importance of adjacent nodes, are computed using the following formula.

\begin{equation}
 e_{i,j} = LeakyReLU(\textbf{\textit{a}}^{T}[\textbf{\textit{W}}x_{i}||\textbf{\textit{W}}x_{j}]),
\end{equation}
where [$\cdot||\cdot$] represents concatenation operation, $x_{i}$ and $x_{j}$ are the vectors of node $i$ and $j$ respectively. $\textbf{\textit{W}}$ is a learnable weight matrix used for linear transformations, and $\textbf{\textit{a}}$ represents a  weight vector. We apply $softmax$ operation to transform these attention coefficients into a probability distribution.
\begin{equation}
 \alpha_{i,j} = \frac{exp(e_{i,j})}{\sum_{k\in N_{i}}exp(e_{i,k})},
\end{equation}
where $\alpha_{i,j}$ is the attention weight obtained from node $j$ to node $i$, and $N_{i}$ denotes the set of neighbors of node $i$. At last, we multiply the features of each adjacent node by its corresponding attention weight, and sum them up to obtain the aggregated representation $x_{i}'$ of node $i$.
\begin{equation}
 x_{i}' = \sigma(\sum_{k\in N_{i}}\alpha_{i,k}\textbf{\textit{W}}x_{k}),
\end{equation}
where $\sigma(\cdot)$ is a customized activation function, $ReLU(x)=max(0,x)$. We utilize GAT layers to construct the encoder.
\begin{equation}
\textbf{H}^{(l+1)}=f(\textbf{H}^{(l)}, \textbf{\textit{W}}^{(l)}),
\end{equation}
where $\textit{f}(\cdot)$ means applying one layer of GAT. $\textbf{H}^{(l)}$ denotes the input of GAT, while $\textbf{H}^{(l+1)}$ is the output, and $\textbf{H}^{(0)}=\textbf{\textit{X}}$.
After applying GAT, we obtain latent representation $\textbf{Z}$. 
\begin{gather}
\textbf{H}^{(1)}=f(\textbf{X}, \textbf{\textit{W}}^{(0)}), \\
\textbf{Z}=\textbf{H}^{(2)}=f(\textbf{H}^{(1)}, \textbf{\textit{W}}^{(1)}).
\end{gather} 
Next, we employ a reconstruction decoder consisting of two layers of GAT. This decoder is based on the latent representation $\textbf{Z}$ to predict the original node attributes.
\begin{gather}
\textbf{H}^{(3)}=f(\textbf{Z}, \textbf{\textit{W}}^{(2)}), \\
\mathbf{\hat{X}}=\textbf{H}^{(4)}=f(\textbf{H}^{(3)}, \textbf{\textit{W}}^{(3)}).
\end{gather}
Finally, we compute the difference between the original matrix $\textbf{X}$ and the reconstructed matrix $\mathbf{\hat{X}}$. The objective function of the model can be formulated as follows,
\begin{equation}
\mathcal{L}=\textbf{R}_{A}=||\textbf{X}-\mathbf{\hat{X}}||^{2}_{F}.
\end{equation}
By minimizing the reconstruction error $\textbf{R}_{A}$, we train the graph auto-encoder to accurately reconstruct the graph features. After a certain number of iterations, we calculate the score for each node $v_i$ using the following formula.
\begin{equation}
score(v_{i})=||x_{i}-\hat{x_{i}}||_{2}.
\end{equation}
Nodes with higher score indicate a greater dissimilarity compared to the majority. Because their patterns deviate significantly from the common cases and hard to be accurately reconstructed. Notably, training the GAT-based graph auto-encoder does not rely on node labels, so \textit{FaaSRCA} is unsupervised.

\subsection{Root Cause Analysis}
Anomaly detection is straightforward in FaaS, such as identifying issues through request success rate and latency. Our focus is on identifying the root causes of these anomalies. We consider requests to serverless application as the object for root cause analysis. Given that the global call graphs of different types can vary significantly in terms of topology and node attributes, to improve the performance of root cause analysis, it is advisable to classify these graphs first.

A serverless application executes specific tasks based on the input parameters provided by the request, following the corresponding call path. Therefore, we can classify requests by analyzing the values of parameters (host, URL, etc.) included in them. Requests with identical parameter values are grouped into the same type. For instance, requests with parameter of  \textit{``http.host=query-orders-for-refresh.default.xxx''} are classified into the category of \textit{``query-orders-for-refresh''}. When a request branches conditionally based on input parameters, we consider different branches to belong to different types. Under this categorization approach, we can easily identify the trace information for a specific serverless application from the serverless platform executing multiple applications. Then, using the metadata (e.g., pod name) extracted from the trace, we can query the observability data for that application from the cluster-level monitoring data collected, without the need to know where its instances are allocated.

After classifying the global call graphs, we locate the root cause nodes within them. We observe the presence of heterogeneity among nodes in the global call graph. Nodes originating from the platform and application side exhibit differences in terms of attributes and reconstruction errors. If we simply rank nodes based on their reconstruction errors and consider high-scoring nodes as root causes, we may overlook the impact of node types. Furthermore, different functions can have various normal patterns, such as different metric values or trace latency. It is not appropriate to define anomalies for all functions using the same criteria. Therefore, considering the global call graph as a heterogeneous graph is crucial.

To solve this problem, we compare the node scores in the global call graph with their corresponding normal ones to measure their deviation from the normal patterns. We begin by classifying the global call graphs obtained during fault-free phase based on their associated request types. Subsequently, we calculate the mean and standard deviation of their node scores, enabling us to obtain the distribution of node scores under normal circumstances. During the faulty phase, we classify graphs into their respective types and align them with the corresponding normal ones. Next, we use \textit{z-score} to calculate the degree of deviation between the scores of each node and the normal ones. \textit{Z-score} is a standardized score that quantifies the deviation of data from the mean using the following formula. $\mu=\frac{1}{N}\sum^{N}_{i=1} x_{i}, \sigma=\sqrt{\frac{1}{N}\sum^{N}_{i=1}(x_{i}-\mu)^{2}}, \textbf{z}_{i} = (x_{i}-\mu)/\sigma$, 
where $N$ is the total number of normal global call graphs, $x_{i}$ is the score of node $i$, $\mu$ is the mean of all scores, $\sigma$ is the standard deviation, and $\textbf{z}_i$ is the \textit{z-score}.

\begin{figure}[t]
    \centering
    \includegraphics[width=0.47\textwidth]{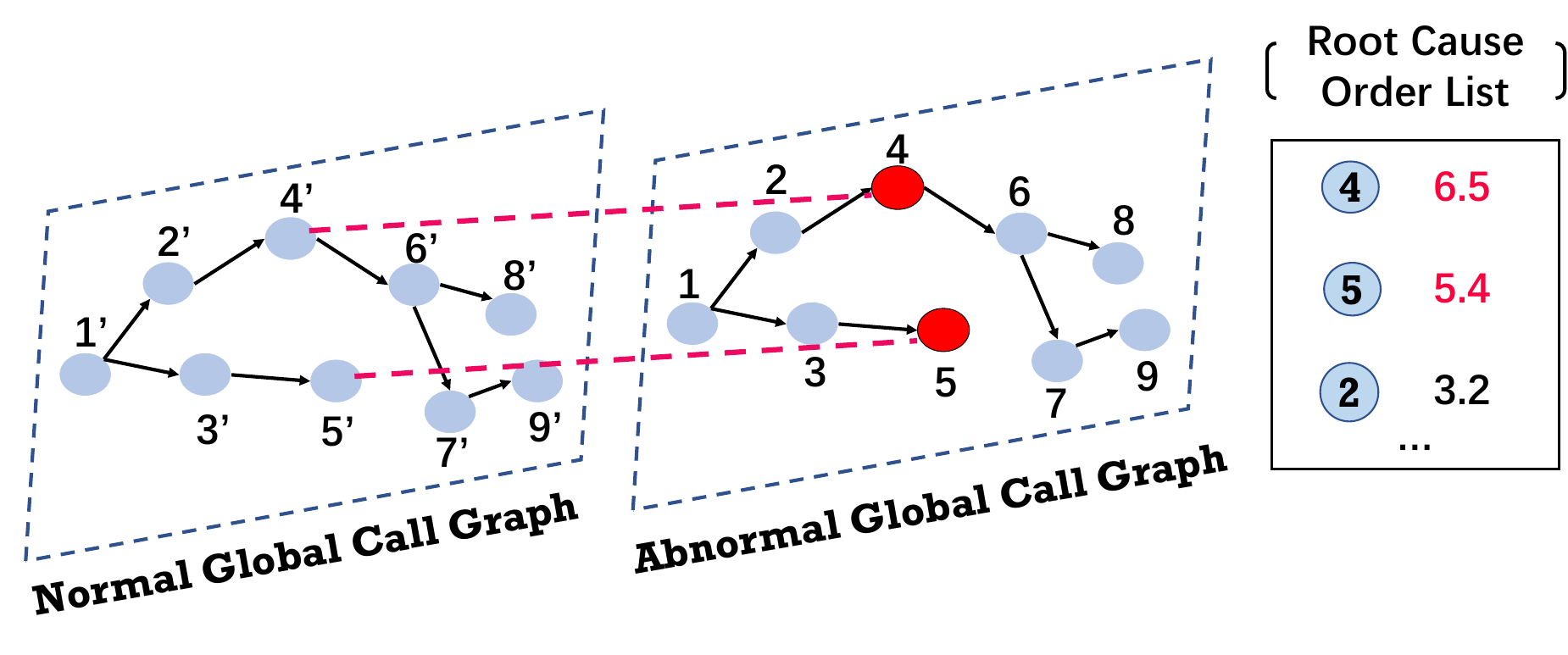}
    \captionsetup{font=normal}
    \caption{Visualization of locating root cause nodes.}
    \vspace{-0.2in}
    \label{fig:root_cause_analysis}
\end{figure}

Finally, we rank the nodes based on their \textit{z-score}, and identify higher-ranked nodes as potential root causes. Our method accurately locates the root cause within heterogeneous global call graphs by evaluating the degree of deviation between each node and its own normal state, without being affected by the diverse attributes of the nodes. Fig.~\ref{fig:root_cause_analysis} provides an example of root cause analysis in \textit{FaaSRCA}, where the root causes are visually highlighted in red for emphasis.

\section{Evaluation}
We address the following research questions (RQs).
\begin{itemize}[itemsep=0pt, topsep=3pt]
\item \textbf{RQ1:} How effective is \textit{FaaSRCA} in full lifecycle root cause analysis compared to existing baseline methods?
\item \textbf{RQ2:} How much does the multi-modal data from the platform and application side contribute to \textit{FaaSRCA}?
\item \textbf{RQ3:} How do different configurations impact the effectiveness of \textit{FaaSRCA}?
\end{itemize}

\subsection{Experiment Setup}
\textbf{Benchmark Datasets.} Our experiments are conducted on two serverless benchmarks, which are widely used in many previous studies \cite{wang2022enhancing, liu2020unsupervised, zhou2019latent, chen2022deep, yu2021microrank, YuCZ19, zhang2022deeptralog}. Serverless TrainTicket \cite{ServerlessTrainTicket2023} is a serverless version modified from the open-source microservice ticket booking system TrainTicket \cite{zhou2018benchmarking}. It includes various types of requests such as query, pay and cancel tickets, totaling 31 distinct functions. ML Workflow \cite{wang2022enhancing} consists of machine learning workflows composed of 9 serverless functions, adapted from publicly available examples of Azure functions. These workflows perform diverse tasks such as images preprocess and classification.


\textbf{System Platform.} 
To manage the deployment and operation of serverless functions, we deploy the serverless benchmark datasets on a Kubernetes cluster with three virtual machines. Each virtual machine has a 4-core 2.70GHz CPU, 8GB of memory, and runs the Ubuntu 18.04 operating system. We utilize the open-source serverless framework, Knative \cite{Knative2023}, to deploy and run the serverless benchmarks. In order to simulate user requests to the system, we implement a load generator. Additionally, we deploy a set of monitoring tools to collect multi-modal data from the serverless applications. We use cAdvisor \cite{cadvisor2023} to monitor the instantaneous metrics of each serverless function instance. The Node Exporter on each node is used to export metrics to Prometheus \cite{Prometheus2023}, then stored in InfluxDB \cite{Influxdb2023}. For logs, we utilize Fluentd \cite{Fluentd2023} and Elasticsearch \cite{Elastic2023} to collect and store them. For traces, we employ the distributed trace system Jaeger \cite{Jaeger2023} to collect them and use Elasticsearch as the storage back end. 

\textbf{Experimental Settings.} 
We use Python 3.8.16, PyTorch 2.0.1, and PyTorch Geometry 2.3.1 to implement \textit{FaaSRCA}. All methods are implemented using widely-used libraries in Python. For the setting of hyper-parameters, the GAT model is with layer number of 4, hidden layer dimension of 32, and batch size of 128. In the experiments, we propose to optimize the loss function with Adam \cite{KingmaB14} algorithm in each epoch, with an initial learning rate of 0.004, and train 100 epochs. 

\textbf{Data Collection.} 
During the fault-free phase, we collect normal data from the system. This is not challenging because well-designed systems with sufficient resources tend to operate in normal state. If anomalies do occur, we can monitor the data generation process to ensure data quality, and handle them using rule-based methods. Even if there are minor abnormal noises present in the normal data, their impact on \textit{FaaSRCA} is limited. Because the proportion of anomalies is typically low, causing insignificant deviations in the normal patterns.

In the faulty phase, to simulate real-world scenarios, we inject faults into the system to generate abnormal data. FaaS lifecycle encompasses both the platform and application sides. On platform side, we use Chaosblade \cite{Chaosblade2023} to inject common issues including network errors and Kubernetes component (kube-scheduler, kubelet, etc.) failures. On application side, we utilize Chaosblade to induce metric failures like CPU utilization spikes by overusing resources. Furthermore, we introduce bugs into function code to simulate code defects, which lead to increased execution time or resource consumption. To evaluate the effectiveness of \textit{FaaSRCA}, we inject a total of 9 typical types of faults, and the detailed list is provided in Table~\ref{tab:fault_type}. When faults are injected into serverless applications, it may cause one or several of their functions to fail.

\begin{table}[h]
  \centering
  \caption{The fault categories injected into the benchmark.}
  \label{tab:fault_type}
  \begin{tabularx}{\linewidth}{>{\hsize=0.5\hsize\linewidth=\hsize}X|
  >{\hsize=1.5\hsize\linewidth=\hsize}Xm{2cm}}
    \hline
    \textbf{Category} & \textbf{Type} \\
    \hline
    \multirow{2}{*}{Platform Fault} & pod failure; replicaset failure; kube-scheduler delay; kubelet delay; network failure\\
    \hline
    Application Fault & code defect; memory stress; CPU contention \\
    \hline
  \end{tabularx}
\end{table}

\begin{table*}[h]
\renewcommand{\arraystretch}{1.4}
  \centering
  \caption{Effectiveness comparison of different approaches for root cause localization.}
  \label{tab:evaluation_metrics}
  \begin{tabular}{m{70pt}<{\centering}m{40pt}<{\centering}m{40pt}<{\centering}m{40pt}<{\centering}m{40pt}<{\centering}|m{40pt}<{\centering}m{40pt}<{\centering}m{40pt}<{\centering}m{40pt}<{\centering}}
    \hline\\[-3.5mm]\hline 
    \multirow{2}{50pt}{\centering \textbf{Approach}} & \multicolumn{4}{c}{TrainTicket}   &\multicolumn{4}{c}{ML Workflow} \\ 
    \cline{2-5}\cline{6-9}
    & \textit{HR@k} & \textit{HR@k+2} & \textit{NDCG@k} & \textit{NDCG@k+2} &  \textit{HR@k} & \textit{HR@k+2} & \textit{NDCG@k} & \textit{NDCG@k+2} \\ 
    \hline\hline
    TBAC & 11.77 & 26.40 & 11.77 & 23.67 &  10.11 & 20.39 & 17.59 & 26.59 \\
    MonitorRank & 8.72 & 27.24 & 8.72 & 21.10 &  11.09 & 21.20 & 13.51 & 21.09 \\
    TraceAnomaly & 22.80  & 25.42 & 25.30  & 27.52  & 33.36 & 34.40 & 51.26 & 50.89\\
    MicroRank & 16.79  & 55.84 & 17.35  & 51.60  & 29.06 & 46.55 & 42.15 & 59.36 \\
    PDiagnose & 48.01  & 68.84 & 49.66  & 67.04  & 54.97 & 70.81 & 63.17 & 69.75\\
    Eadro & 73.34 & 89.70 & 73.37 & 83.40  & 67.23 & 89.49 & 67.99 & 80.95\\
    \hline
    Direct-FaaSRCA & 68.74 & 82.28 & 69.25 & 79.15  & 58.48 & 65.86 & 61.66 & 65.04\\
    FaaSRCA & \textbf{92.50} & \textbf{96.33} & \textbf{93.48} & \textbf{95.98}  &\textbf{90.57} & \textbf{95.97} & \textbf{95.76} & \textbf{97.32}\\
    \hline\\[-3.5mm]\hline
  \end{tabular}
\end{table*}

Through the above methods, we collect a total of 154,307 global call graphs for Serverless TrainTicket. Out of these, we use 10,000 graphs for model training. Among the remaining graphs, 74,135 are normal data generated during the fault-free phase, which are used to establish the normal patterns. The other 70,172 graphs are abnormal data, which serve as the test data. For ML Workflow, we collect 58,354 global call graphs, with 5,000 graphs used for training. And there are 21,045 normal graphs and 32,309 abnormal graphs remaining. In real-world serverless systems, it is relatively straightforward to collect this amount of information for commonly used serverless applications. For applications with lower invocation frequencies, we only need a sufficient number of global call graphs that represent their normal patterns.

\textbf{Metrics.} 
We introduce hit ratio (HR) and normalized discounted cumulative gain (NDCG) at the top-k as the evaluation metrics of root cause localization. The number of root causes may vary across different graphs, so we determine it as the initial top-k value. For instance, if there is a single root cause, k is set to 1, and HR@k refers to HR@1, which examines the top 1 result in the predicted list. In this scenario, HR@k+2 corresponds to HR@3, and so on.

HR@k indicates whether the predicted top-k list contains the root cause. $\textit{HR@k} = \frac{1}{N}\sum^{N}_{i=1}(s^{i}\in S^{i}_{[1:k]})$, where $S^{i}_{[1:k]}$ represents the predicted top-k list for the graph $i$, $s^{i}$ denotes the actual root causes, and N is the number of abnormal graphs. $NDCG@k$ measures the extent to which the root causes appear at higher positions in the candidate list. $\textit{DCG@k} = \frac{1}{N}\sum^{N}_{i=1}(\sum_{i=1}^{k} \frac{2^{rel_i} - 1}{\log_2(i+1)}),
 \textit{IDCG@k} = \frac{1}{N}\sum^{N}_{i=1}(\sum_{i=1}^{k} \frac{2^{rel_i^*} - 1}{\log_2(i+1)}),
 \textit{NDCG@k} = \frac{DCG@k}{IDCG@k}
$, $rel_i$ represents whether the position $i$ is a root cause, while $rel_i^*$ represents whether the position $i$ sorted in descending order of relevance scores is a root cause. 

\subsection{Baselines}
\begin{itemize}
\item \textbf{TBAC} \cite{MarwedeRHH09} ranks root causes by aggregating the Pearson correlation coefficients of the service dependency graph.
\item \textbf{MonitorRank} \cite{KimSS13} employs a customized random walks with self edges and back edges to identify root causes.
\item \textbf{TraceAnomaly} \cite{liu2020unsupervised} employs a posterior flow based variational auto-encoder (VAE) to learn the normal patterns of traces offline. Traces with lower abnormal scores are considered as anomalies, and undergo root cause analysis.
\item \textbf{MicroRank} \cite{yu2021microrank} performs statistical analysis on trace duration and considers traces deviating from normal pattern as anomalies. Personalized PageRank and spectrum methods are then combined to locate the root cause.  
\item \textbf{PDiagnose} \cite{hou2021diagnosing} takes multi-modal data as input and converts them into time series. It employs vote based problem localization strategy to determine the root cause.
\item \textbf{Eadro} \cite{Eadro23} proposes a supervised end-to-end framework based on multi-source data integration for anomaly detection and root cause localization. It utilizes shared knowledge of these two phases via multi-task learning.
\item \textbf{Direct-FaaSRCA} is a variant of \textit{FaaSRCA} that directly determines the root causes based on the ranking of reconstruction errors in the graph.
\end{itemize}

These above RCA baselines are incapable of accurately modeling serverless applications, and overlook the root causes within platform side. Consequently, we make corresponding modifications to adapt them for use in serverless scenarios.

\subsection{\textbf{RQ1: Effectiveness in Root Cause Analysis.}}

To perform a thorough evaluation of \textit{FaaSRCA}, we compare its performance with baseline methods. Table~\ref{tab:evaluation_metrics} shows the evaluation results of different methods on two datasets. 

TBAC, MonitorRank, TraceAnomaly, and MicroRank are trace-based methods that utilize various techniques such as statistical models, heuristic methods, spectral algorithms, and more to identify the root causes of faults. However, their performance is relatively low, with average HR@k values of only 10.94\%, 9.91\%, 28.08\%, and 22.93\% respectively. This is because they overlook valuable information contained in logs and metrics. Anomalies within logs and metrics may not necessarily manifest as anomalies in traces, making it difficult for trace-based only root cause analysis methods to effectively troubleshoot them. On the other hand, unlike microservices where a single service may contain multiple tasks, each serverless function typically handles smaller granular tasks. This leads to a decrease in the diversity of traces that pass through the same function in serverless applications, and reduces coverage of trace types that encompass functions. Consequently, the effectiveness of spectral methods are diminished.

Compared to single modal methods, PDiagnose leverages multi-modal data and has better performance. However, its HR@k and NDCG@k, averaging at 51.49\% and 56.42\% respectively, still fall short of expectations. This can be attributed to following reasons. First, PDiagnose processes logs by counting the occurrence of error keywords and ignores their semantic information. Second, it just utilizes the latency of traces and disregards their topology, relying solely on a voting mechanism for root cause analysis. In addition, PDiagnose relies on manual parameter settings and thresholds, which are based on expertise or learned from historical data. These rule-based designs also affect the performance of model.

Eadro effectively learns meaningful patterns from multi-modal data and integrates them through graph structures. But compared to \textit{FaaSRCA}, it exhibits a reduction of 21.25\% in HR@k and 23.94\% in NDCG@k. This discrepancy can be attributed to its inability to handle the heterogeneity of nodes within the graph (also the case with \textit{Direct-FaaSRCA}). Treating nodes from different sides as isomorphic disregards their distinct attributes, making it challenging to identify the root cause. In \textbf{RQ3}, we conduct experiments to discuss the implicit consideration of heterogeneous graphs by \textit{FaaSRCA}. Besides, as previously mentioned in \S~\ref{sec:motivation}, the lack of sequence continuous data degrades Eadro's performance. Moreover, unlike unsupervised \textit{FaaSRCA}, Eadro is a supervised method and requires labeled data, making it laborious and time-consuming.

The principle behind \textit{Direct-FaaSRCA} is that, the more challenging it is to reconstruct a node, the greater the reconstruction error, and the more likely it is to be identified as the root cause. However, due to the heterogeneity between platform and application nodes, its performance is much poorer than \textit{FaaSRCA}. This underscores the necessity of our heterogeneous graph approach. \textit{FaaSRCA} compares the reconstructed graph to the normal one, and identifies nodes with the largest error deviation as root causes.

Across all experiments results, \textit{FaaSRCA} achieves the best effectiveness, attains an average HR@k of 91.54\% and NDCG@k of 94.62\%. Compared to the baseline methods, \textit{FaaSRCA} demonstrates a higher HR@k ranging from 21.25\% to 81.63\% and NDCG@k ranging from 23.94\% to 83.51\%. These results indicate the effectiveness of \textit{FaaSRCA} in accurately locating the root cause under various fault scenarios. 

\subsection{\textbf{RQ2: Contributions of Multi-modal Data From Platform and Application Side.}}
\label{subsec:exp_multi_modal}
To explore the contributions of different data sources, we conduct ablation study. The variants of \textit{FaaSRCA}, denoted as \textit{FaaSRCA w/o X}, are executed by excluding specific data source. \textit{FaaSRCA w/o M} drops metrics, \textit{FaaSRCA w/o L} removes logs, \textit{FaaSRCA w/o T} eliminates traces latency. 

Furthermore, to confirm the contribution of platform side data and validate the scope of root causes that \textit{FaaSRCA} can cover in serverless applications, we design the following experiments. \textit{FaaSRCA w/o P} indicates not considering platform side data, and only locates root causes on the application side. Additionally, we compute the effectiveness of \textit{FaaSRCA} in separately localizing root causes on both sides, denoted as \textit{FaaSRCA on P} and \textit{FaaSRCA on A} respectively.

\begin{table}[t]
\renewcommand{\arraystretch}{1.4}
  \centering
  \caption{Experimental results of ablation study.}
  \label{tab:ablation_study}
\begin{tabular}{m{60pt}<{\centering}m{25pt}<{\centering}m{35pt}<{\centering}|m{25pt}<{\centering}m{35pt}<{\centering}}
\hline\\[-3.5mm]\hline
\multirow{2}{60pt}{\centering \textbf{Ablation}} & \multicolumn{2}{c}{TrainTicket} &  \multicolumn{2}{c}{ML Workflow} \\
    \cline{2-3}\cline{4-5}
    & \textit{HR@k} & \textit{NDCG@k} &   \textit{HR@k} & \textit{NDCG@k} \\
    \hline\hline
FaaSRCA w/o M  & 67.84 & 69.34 &  59.65 & 66.58\\
FaaSRCA w/o L  & 63.09 & 64.99 &  69.64 & 72.05\\
FaaSRCA w/o T  & 58.61 & 60.08 &  57.98 & 62.62\\
FaaSRCA w/o P  & 62.84 & 67.89 &  41.84 & 63.23\\ 
FaaSRCA & \textbf{92.50} & \textbf{93.48} & \textbf{90.57} & \textbf{95.76}\\
\hline
FaaSRCA on P & 93.81 &94.80 &  90.59 &95.79\\
FaaSRCA on A & 89.62 & 90.56 & 90.37 & 95.43\\

\hline\\[-3.5mm]\hline
\end{tabular}
\end{table}

The results of the ablation experiments are shown in Table~\ref{tab:ablation_study}. It is evident that \textit{FaaSRCA} achieves the best performance, and each data modality contributes to its effectiveness. The performance of \textit{FaaSRCA w/o M}, \textit{FaaSRCA w/o L} and \textit{FaaSRCA w/o T} sharply declines, as metrics, logs, and traces offer valuable information about serverless functions from different perspectives, aiding root cause identification.

\textit{FaaSRCA w/o P} solely relies on application side data for root cause analysis. In this case, it may fail to detect faults occurring on the platform side. As a result, it struggles to accurately capture the root cause, resulting in a decrease in its performance, which highlights the importance of considering both platform and application side factors. Furthermore, we observe that \textit{FaaSRCA} performs similarly on the platform and application side, indicating its ability to encompass the entire lifecycle of serverless functions. \textit{FaaSRCA} can perform root cause localization with a finer granularity throughout the entire lifecycle of a function, whereas previous methods were limited to instance-level granularity. In general, the success of \textit{FaaSRCA} can be attributed to its fully integration and utilization of multi-modal data from both sides.

\subsection{\textbf{RQ3: Impact of Configurations.}}
(1) \textit{\textbf{Configurations of GAT.}} To evaluate the impact of GAT configuration parameters on \textit{FaaSRCA} performance, we conduct several experiments on Serverless TrainTicket dataset. 

\begin{figure}[h]
    \centering
    \includegraphics[width=0.5\textwidth]{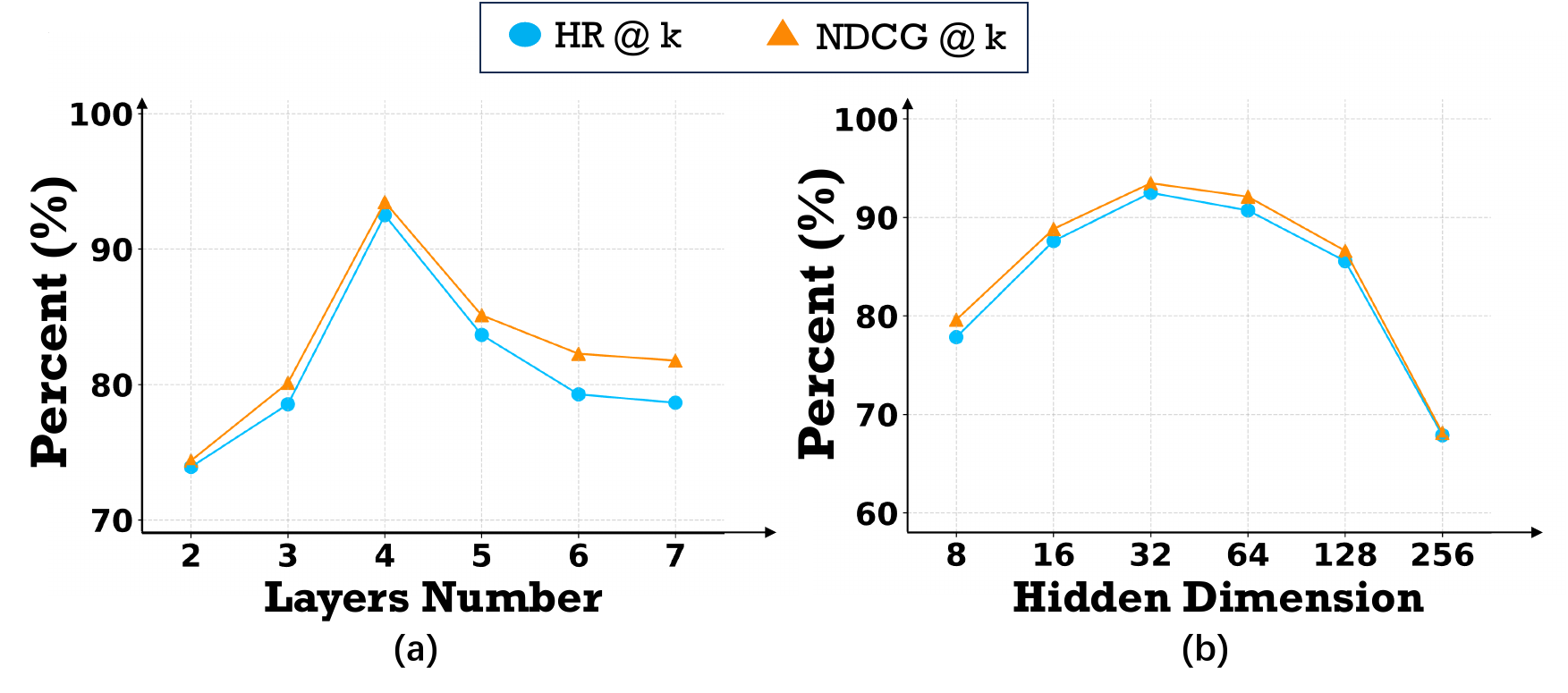}
    \captionsetup{font=normal}
    \caption{(a) Impact of GAT layer number. (b) Impact of GAT hidden layer dimension.}
    \vspace{-0.2in}
    \label{fig:configuration}
\end{figure}

The number of GAT layers determines the depth of the network architecture through which the data passes during the encoding and decoding process. Fig.~\ref{fig:configuration}(a) shows its impact on \textit{FaaSRCA}. It can be seen that \textit{FaaSRCA} achieves the best results when the GAT layer is set to 4, while fewer or more layers result in lower performance. When there are fewer layers, the transmission of information between different nodes may be insufficient, which limits model's ability to learn and represent the complex relationships within the graph. However, including more layers can easily lead to over-smoothing of the model, making it difficult to distinguish features.

Fig.~\ref{fig:configuration}(b) illustrates the influence of the GAT hidden layer dimension. The choice of an appropriate dimension is crucial in achieving optimal performance. Lower dimensions may restrict model's capacity to extract meaningful features from the data, resulting in information loss and underfitting. Conversely, higher dimensions can empower the model to learn more intricate features, but it may lead to overfitting, increased model complexity, and longer computational time. 

(2) \textit{\textbf{Configurations of graph neural network basis.}} To verify the model's ability to learn node representations, we select different graph neural network as the basis of Graph Auto Encoder for comparison, including GCN, GAT, and Simple-HGN. Simple-HGN \cite{lv2021we} is a heterogeneous graph neural network that inherits from the simple homogeneous GAT. It extends the original graph attention mechanism by incorporating edge type information into attention computation. $e_{i,j} = LeakyReLU(\textbf{\textit{a}}^{T}[\textbf{\textit{W}}x_{i}||\textbf{\textit{W}}x_{j}||\textbf{\textit{W}}_{r}r_{\psi(i,j)}])$, where $\psi(i,j)$  represents the type of edge between node $i$ and $j$, $\textbf{\textit{W}}_{r}$ is a learnable matrix to transform type embedding. 

\begin{figure}[t]
    \centering
    \includegraphics[width=0.5\textwidth]{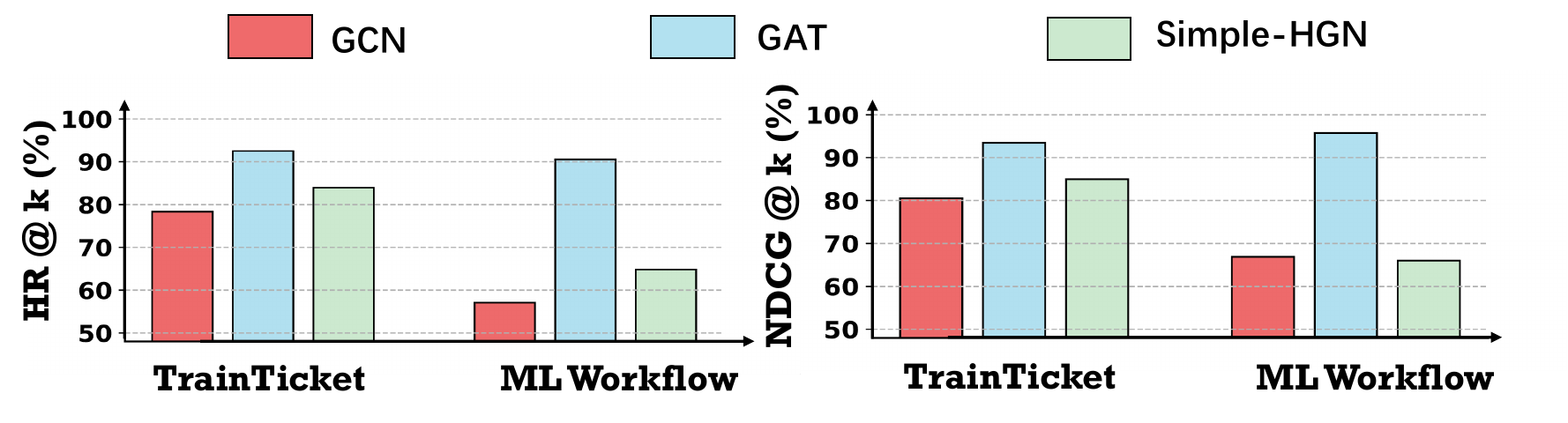}
    \captionsetup{font=normal}
    \caption{Results of different graph neural network basis.}
    \vspace{-0.20in}
    \label{fig:backbone}
\end{figure}

The experimental results depicted in Fig.~\ref{fig:backbone} demonstrate the performance of different base models, with optimal parameters chosen for each. Among all, GAT achieves the best performance. Compared to GAT, GCN shows a average decrease of 23.84\% in HR@k and 20.91\% in NDCG@k. This can be attributed to the simple aggregation operation of GCN, which may result in excessive smoothing, limiting its ability to extract informative features. In contrast, GAT utilizes an adaptive attention mechanism that dynamically learns the importance of each node and its neighboring nodes. This enables GAT to more effectively capture and utilize the graph information.

Despite introducing edge type embedding on the foundation of GAT to handle heterogeneous graphs, Simple-HGN fails to outperform GAT, instead, exhibits lower performance. Specifically, Simple-HGN experiences an average decrease of 17.15\% in HR@k and 19.64\% in NDCG@k compared to GAT. This may be because the heterogeneity of node features in global call graph already implies different types. \textit{FaaSRCA} implicitly addresses this issue by comparing nodes with their normal patterns. However, the incorporation of additional edge type embedding and complex attention mechanism calculations in Simple-HGN, can interfere with the learning representation of nodes, leading to a decline in its performance.

\subsection{Efficiency of \textit{FaaSRCA}}
To demonstrate the applicability of \textit{FaaSRCA}, we evaluate its performance efficiency. The required time of model training depends on epochs, neural network depth, and hidden layer size. In our experiments, training on 5,000 graphs takes 145 seconds, averaging 0.029 per graph. The training process is a one-time computation, reusable without retraining.

Then we calculate the time for \textit{FaaSRCA} to identify root causes during test inference, and compare its efficiency against other methods. In ML Workflow dataset, it takes an average of 175 seconds on the 21,045 graphs. For the 70,172 graphs in the Serverless TrainTicket, \textit{FaaSRCA} takes 597 seconds on average to identify the root causes of failures. In comparison, MicroRank and PDiagnose take an average of 578 and 491 seconds respectively to diagnose the problems, but their root cause analysis accuracy is much lower than \textit{FaaSRCA}. Overall, \textit{FaaSRCA} can identify the root cause for each graph within just 8 milliseconds, indicating its ability to perform root cause analysis in a timely and efficient manner.

\subsection{Discussion}
We identify the following limitations of \textit{FaaSRCA}. Firstly, real-world serverless systems may lack the capability to gather multi-modal data from platform and application side. To utilize \textit{FaaSRCA} effectively, it is essential to enable them to collect such observability data. However, configuring and using these commonly-used monitoring tools is a straightforward task. For existing real-world serverless platforms (e.g., AWS Lambda and Google Cloud Functions), developers only need to integrate the relevant monitoring tools, without any other additional work, to be compatible with and apply our proposed method. Our solution has been shown in \S\ref{subsec:data_preparation}.

Secondly, \textit{FaaSRCA} is unable to identify root causes of failures that go undetected as anomalies, such as abnormal platform infrastructure metrics or Byzantine faults that do not lead to changes in the observability data. However, compared to single modal approaches, \textit{FaaSRCA} has combined multiple data sources, allowing it to cover a wider range of failure types and achieve better root cause analysis.

\section{Related Work}
Metric based RCA methods commonly diagnose problems by mining the relationships between different metrics \cite{lin2018microscope, WangWJHWKX21, wang2018cloudranger, Wu2021Microdiag, wu2020microrca}. CloudRanger \cite{wang2018cloudranger} and Microscope \cite{lin2018microscope} utilize the PC algorithm to construct causal relationship graphs on anomaly metrics. Based on the correlation of different metrics on the causal path, they employ depth-first search and random walk algorithms respectively, to identify the root cause in the graph. MicroRCA \cite{wu2020microrca} uses microservice topology to construct an attribute graph, extracts anomalous subgraphs from it, and uses personalized PageRank method to locate root cause.

Log based RCA methods generally locate root cause by comparing the frequent patterns between normal and abnormal logs. LogCluster \cite{lin2016log} employs log clustering techniques to extract patterns from logs, identifies previously unseen log sequences. SBLD \cite{rosenberg2020spectrum} applies spectral algorithms \cite{jones2002visualization}, locates the root cause by analyzing the coverage range of logs. 

Trace based RCA methods employ different techniques to distinguish between normal and abnormal traces. MicroRank \cite{yu2021microrank} identifies anomalies by detecting traces that deviate from normal patterns, and utilizes spectral algorithms to pinpoint the root cause. TraceRank \cite{yu2021tracerank} encodes traces into vectors by considering the services they traverse, applies hierarchical clustering and k-means algorithm to detect abnormal traces.

Recently, to achieve more accurate analysis than single modal methods, there has been an growing interest in integrating multi-modal observable data from diverse sources \cite{chen2022deep, hou2021diagnosing, Twin23, Eadro23, zhang2022deeptralog, zhang2021cloudrca, zhao2021identifying}. DeepTraLog \cite{zhang2022deeptralog}
utilizes a unified graph representation to capture the complex structure of the traces and logs embedded within them. It then applies Deep SVDD model, based on GGNN, for anomaly detection purposes. SCWarn \cite{zhao2021identifying} employs LSTM model to capture temporal dependencies in each time series, and utilizes multi-modal learning to detect bad software changes. Eadro \cite{Eadro23} presents an end-to-end framework that combines anomaly detection and root cause analysis using multi-source data. 

However, some of these multi-modal methods are supervised \cite{Eadro23} or semi-supervised \cite{Twin23}, which necessitates labeled training data. Acquiring such data can be expensive and impractical in real-world scenarios. Besides, certain studies \cite{zhang2022deeptralog, zhao2021identifying, Twin23} primarily focus on anomaly detection. They learn representation vectors of the entire system and utilize classification techniques to identify anomalies. But they do not address the issue of fine-grained root cause analysis. What's more, these multi-modal root cause analysis methods are mainly used in microservice systems and don't take into account the characteristics of serverless. As we highlight in the section \S~\ref{sec:motivation}, these methods are not well-suited for serverless applications with short lifetimes and high dynamism. Furthermore, they only focus on the running stage of functions while neglecting other stages, thus lacking consideration for platform side data. Borges et.al. \cite{borges2021faaster} and  
Satapathy et.al. \cite{SatapathyTCC23} explore distributed tracing to improve observability in serverless applications. However, they lack quantitative fault localization experiments and root cause analysis implementation. This inspires us to propose \textit{FaaSRCA}, a full lifecycle root cause analysis method tailored for serverless applications.

\section{Conclusion}
In this paper, we propose \textit{FaaSRCA}, a method for root cause analysis of serverless applications throughout their full lifecycle. \textit{FaaSRCA} uses \textit{Global Call Graph} to integrate multi-modal observability data generated from the platform and application side. By employing different modal methods, we capture the features of individual nodes. Then we train a GAT based graph auto-encoder and calculate the reconstruction score for each node. We measure the deviation between the scores of each node and their corresponding normal patterns. Finally, we rank the deviation to locate root causes. We conduct extensive experimental evaluations on two serverless datasets, \textit{FaaSRCA} achieves HR@k of 91.54\% and NDCG@k of 94.62\% on average, exceeding all other baseline methods. In future work, we will attempt to obtain richer information from the platform side, and apply \textit{FaaSRCA} to more serverless systems to further evaluate its effectiveness.

\section{Acknowledgment}
The research is supported by the National Key Research and Development Program of China (No. 2019YFB1804002), the National Natural Science Foundation of China (No. 62272495), the Guangdong Basic and Applied Basic Research Foundation (No.2023B1515020054). This work was supported by Ant Group. The corresponding author is Pengfei Chen.

\bibliographystyle{IEEEtran}

\end{document}